# Unconventional cross sections in zinc phosphide nanowires grown using exclusively earth-abundant components

Simon Escobar Steinvall[1*], Hampus Thulin[1], Nico Kawashima[2,3], Francesco Salutari[4], Jonas Johansson[5], Aidas Urbonavicius[1], Sebastian Lehmann[1], Maria Chiara Spadaro[4,6], Jordi Arbiol[4,7], Silvana Botti[2], Kimberly A. Dick[1]

1. Center for Analysis and Synthesis and NanoLund, Lund University, Box 124, 221 00 Lund, Sweden
2. Research Center Future Energy Materials and Systems and Interdisciplinary Centre for Advanced Materials Simulation, Faculty of Physics and Astronomy, Ruhr University Bochum, Universitätsstraße 150, 44801 Bochum, Germany
3. Institute of Condensed Matter Theory and Optics, Friedrich-Schiller-Universität Jena, Max-Wien-Platz 1, 07743 Jena, Germany
4. Catalan Institute of Nanoscience and Nanotechnology (ICN2), CSIC and BIST, Campus UAB, Bellaterra, 08193 Barcelona, Catalonia, Spain
5. Division of Solid State Physics and NanoLund, Lund University, Box 124, 221 00 Lund, Sweden
6. Department of Physics and Astronomy "Ettore Majorana", University of Catania and CNR-IMM Via S. Sofia 64, 95123 Catania, Italy
7. ICREA, Pg. Lluís Companys 23, 08010 Barcelona, Catalonia, Spain
*Corresponding Author: simon.escobar_steinvall@chem.lu.se

## Abstract

To enable lightweight and flexible solar cell applications it is imperative to develop direct bandgap absorber materials. Moreover, to enhance the potential sustainability impact of the technologies there is a drive to base the devices on earth-abundant and readily available elements. Herein, we report on the epitaxial growth of $Zn_3P_2$ nanowires using exclusively earth-abundant components, using Sn as the nanowire catalyst and Si (111) as the substrate. We observe that the nanowires exhibit a triangular cross section at lower temperatures, a pseudo-pentagonal cross section at intermediate temperatures, and a hexagonal cross section in a twin plane superlattice configuration at high temperatures and high V/II ratios. At low temperatures, the surface facets are constricted into a metastable configuration, yielding the triangular morphology due to the symmetry of the substrate, while intermediate temperatures facilitate the formation of a pseudo-pentagonal morphology with lower surface to volume ratio. The twin plane superlattice structure can only be observed at conditions that facilitate the incorporation of Sn into $Zn_3P_2$, which is needed to form heterotwins in the tetragonal structure, namely at high temperatures and high phosphine partial pressures. These findings show a clear pathway to use $Zn_3P_2$ nanowires in sustainable solar energy harvesting using exclusively earth-abundant components, as well as opening up a novel route of fabricating quantum wells inside nanowires using heterotwins.

## Introduction

Diversification of the materials used in the solar energy sector is important to assure its resilience and its potential impact in the sustainable energy transition.[1] There have been a great number of earth-abundant semiconductors proposed for use in photovoltaics based on their bandgap to address this issue.[2,3] However, practical limitations due to their processing or compatibility with other materials can result in significant defect formation within the materials or at interfaces when fabricating devices.[4] The uncontrolled defects have severely limited these materials, with current performance levels being well below their respective theoretical conversion efficiency limits.[2,5] Consequently, significant research effort is being dedicated to developing alternative approaches to overcome the limitations set by conventional processing routes.

One approach that has been demonstrated to reduce interface defect formation in semiconductor heterostructures is nanoscale epitaxy.[4,6–8] Compound semiconductor nanowires in particular have been demonstrated as a diverse platform for combining mismatched lattice materials[6–11], controlling the crystal structure[12–15], and achieving metastable compositions[16,17]. A majority of nanowire research has focused on conventional semiconductors, such as III-V, II-VI and group IV semiconductors. However, the structural benefits of the nanowire morphology can greatly enhance the applicability of emerging compound semiconductors as well. In addition, the introduction of wavelength-scale optoelectronic properties when working with nano-dimensional materials are a powerful tool in further optimising photovoltaics.[18–25]

Zinc phosphide ($Zn_3P_2$) is an earth-abundant compound semiconductor with optoelectronic properties suitable for an absorber in single-junction photovoltaics.[26–30] However, the fabrication of $Zn_3P_2$-based devices is complicated by i) its lattice parameters and coefficient of thermal expansion not matching with any suitable partner material for heterojunctions[31–35] and ii) lack of controlled n-type doping for homojunctions[36–38]. Nanoscale epitaxy of $Zn_3P_2$ has recently been demonstrated as a promising approach to overcome the first limitation and enable heterojunction formation and control the defect formation.[4,11,23,39–41] While the results have been promising with regards to material properties, the process has relied on indium (In) in the substrate (InP) and as the catalyst for vapour-liquid-solid (VLS) growth, although the catalyst contribution to the In usage is relatively minor. In is a scarce element that is generally produced as a byproduct from the production of other materials, mainly Zn, and as such has an inelastic supply-demand relationship and there is a high likelihood of future supply shortages.[42,43] The use of In in the growth effectively negates any earth-abundance and sustainability gains of $Zn_3P_2$ based growth. Tin (Sn) has been demonstrated as a more earth-abundant catalyst material for VLS growth of $Zn_3P_2$[44], though there has been no reported alternatives with regards to a more earth-abundant and sustainable substrate.

In this article, we report the epitaxial growth of $Zn_3P_2$ nanowires using exclusively earth-abundant components. We use silicon (Si) (111) substrates for the epitaxial growth of $Zn_3P_2$ nanowires by metalorganic chemical vapour deposition (MOCVD) with Sn as a VLS catalyst. Furthermore, we produced crystal morphologies not previously reported, including triangular, pseudo-pentagonal and hexagonal cross-sectional nanowires, which can be tuned through the control of temperature and relative V/II ratio during growth. Using a combination of density functional theory (DFT) and scanning transmission electron microscopy (STEM) we can attribute the cross-sectional morphology to metastable facet formation due to epitaxial constraints. Through this work we demonstrate the use of nanoscale epitaxy to enable more sustainable process routes of emerging compound semiconductors and its use in producing metastable crystal morphologies.

## Results and discussion

### Catalyst particle synthesis

To grow $Zn_3P_2$ nanowires by the VLS method the first step involves the deposition of liquid metal catalyst particles. The catalysts enable the growth of nanowires by acting as a sink for the precursors and selectively precipitating the compound semiconductor at the liquid-solid interface upon supersaturation.[45] Sn catalyst particles were deposited in-situ in the MOCVD on Si(111) substrates, which had been cleaned and passivated using buffered HF, using tetrakis(dimethylamino)tin (TDMASn) as the precursor. We observe a high particle density for deposition temperatures ranging from 570 °C to 630 °C. SEM images of particles deposited at different temperatures under identical TDMASn partial pressures and deposition times are shown in Figure 1a. From these images we could extract the particle areas, which in turn were converted into diameters and plotted in Figure 1b. For subsequent growth experiments we used Sn nanoparticles deposited at 630 °C, which yielded average diameters of 45 nm ± 14 nm.

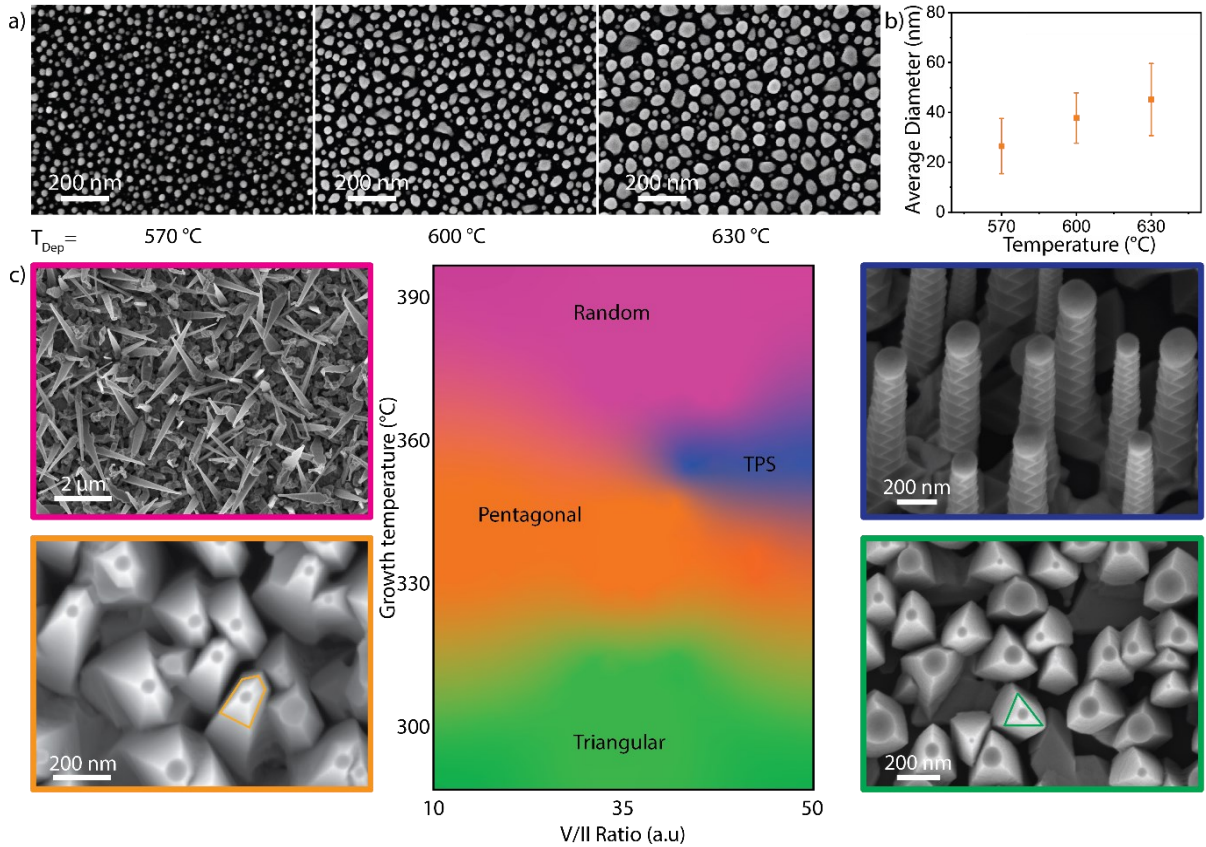

**Figure 1.** (a) Top view SEM images of Sn nanoparticles on Si(111) deposited at varying susceptor temperatures for a deposition time of 20 minutes. (b) is a plot of particle size and distribution as a function of temperature. (c) SEM images showing (purple) randomly oriented nanowires, (orange) pseudo-pentagonal nanowires, (blue) TPS nanowires (imaged at 15° tilt to show their characteristic zigzag shape), and (green) triangular nanowires, as well as the corresponding V/II ratio vs growth temperature plot showing the parameter space for each morphology as indicated by the colour.

**Combinatorial growth study**

To grow $Zn_3P_2$ nanowires we subsequently changed the temperature from the Sn particle deposition temperature to the growth temperature, which was varied between 276 °C and 396 °C. At the growth temperature we then performed a 5-minute Zn pre-deposition by only flowing diethyl zinc (DEZn), similar to what was done for In-catalysed $Zn_3P_2$ nanowires in Ref [11], before turning on the phosphine ($PH_3$) to commence the nanowire growth. By varying the $PH_3$ partial pressure while keeping the DEZn partial pressure constant at $9.72 \times 10^{-1}$ Pa for different temperatures we could then map out the VLS parameter space for Sn-catalysed $Zn_3P_2$ nanowires on Si (111), shown in Figure 1c and SI Figures 1-4.

For ordered nanowire growth we observed three main morphologies in the SEM images, in addition to randomly oriented nanowires under certain conditions. At high temperatures (>356 °C) and V/II ratios the nanowires exhibited the twin plane superlattice structure, which was previously reported for non-epitaxial Sn-catalysed $Zn_3P_2$ nanowires.[44] However, for lower V/II ratios and lower temperatures we start observing alternative morphologies. First, at high temperatures and lower V/II ratios, as well as intermediate temperatures for all V/II ratios, we observe the emergence of an irregular five faceted cross section, i.e. a pseudo-pentagonal cross section. If the temperature is lowered even further, the number of cross-sectional facets goes down to three, forming a triangular cross section. For the highest temperature in the range explored (396 °C) we stop observing any epitaxial ordering or control of the nanowire growth direction or morphology, while at the lowest temperature explored (276 °C) the nanowires retained a triangular cross section but the growth rate started to decrease. The origin of the different cross sections is discussed in more detail in the next section.

The radial overgrowth of the nanowire side facets also shows a temperature and V/II ratio dependence, as can be observed in the SEM images in SI Figures 1-4. For growth at 356 °C we did not observe any significant radial overgrowth, irrespective of V/II ratio. However, for the pseudo-pentagonal nanowires grown at 336 °C we observe a V/II ratio dependence of the radial overgrowth with higher V/II ratios resulting in higher rates of radial overgrowth. There is a facet dependence in the overgrowth, examined in more detailed when discussing surface energies below. Finally, for the triangular nanowires grown at 306 °C we observe significant radial overgrowth irrespective of V/II ratio. The trends in overgrowth can be explained through the surface diffusion of Zn. The mobility of Zn atoms on the surface will initially increase with temperature, increasing the likelihood for Zn atoms deposited on away from the Sn particles to be able to reach them and contribute to VLS growth as their diffusion length increases. At intermediate temperatures the V/II ratio dependence further supports the Zn diffusion length dependence of the radial overgrowth. The group V element will have negligible surface diffusion during nanowire growth.[46,47] Therefore, increasing the V/II ratio at constant Zn partial pressure will decrease the Zn diffusion length and consequently promote radial overgrowth. In addition, the high vapour pressure of Zn may also result in the Zn atoms re-evaporating instead of contributing to the radial overgrowth at a higher rate.

**Cross-sectional analysis**

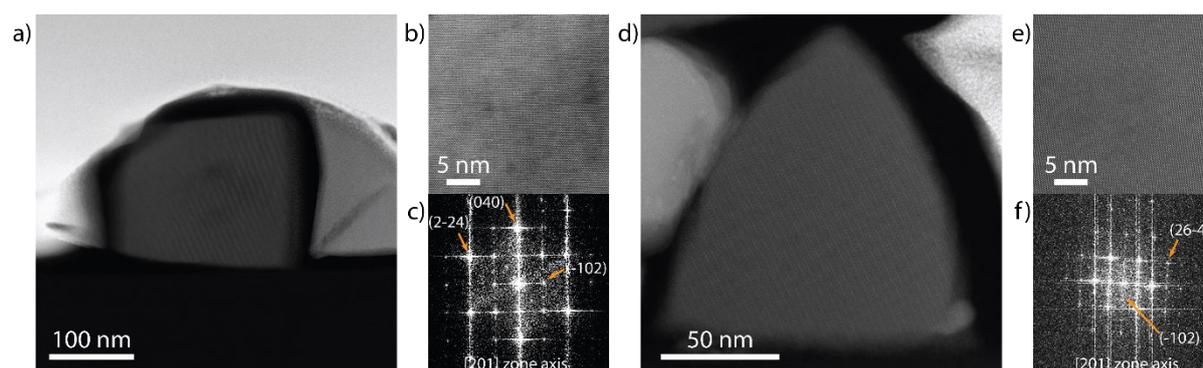

**Figure 2.** (a) Low-magnification AC-HAADF STEM image of a FIB cross section from a pseudo-pentagonal nanowire with corresponding (b) high-magnification image and (c) Fast-Fourier Transform (FFT) power spectrum indicating a [201] zone axis and highlighting the spots related to the surfaces of the nanowire. (d) Low-magnification AC-HAADF STEM image of a FIB cross section from a triangular nanowire with corresponding (e) high-magnification image and (f) FFT power spectrum indicating a [201] zone axis and highlighting the spots related to the surfaces of the nanowire.

Pentagonal symmetries are not commonly observed in crystals, so to ascertain the origin of the cross sections we first investigated the presence of any core-defects, such as penta-twins or voids that can give rise to pentagonal morphologies.[48,49] Using focused ion beam (FIB) fabrication, we prepared electron transparent cross-sectional lamellae of triangular and pseudo-pentagonal nanowires that were analysed using high-angle annular dark field aberration-corrected scanning transmission electron microscopy (HAADF AC-STEM), as shown in Figure 2.

For both the triangular and pseudo-pentagonal nanowires we do not observe any penta-twin or other defects that could potentially explain the morphology. Instead we observe $Zn_3P_2$ single crystals along a <201> zone axis. In $Zn_3P_2$'s tetragonal unit cell, this indicates that the nanowires are growing through the stacking of (101) planes.[11,39] For the triangular nanowires we could thus determine that the side-facets were made up of two {132} facets and one {102} facet.

For the pseudo-pentagonal case we did not observe the {132} facets, instead we observed a combination of {100} facets and {112} facets. However, during radial overgrowth there seems to be significant amount of micro-faceting and deviation from these facets, as previously observed for MBE grown $Zn_3P_2$ nanowires with {100} facets.[39]

**DFT and facet formation analysis**

As we could not observe any crystallographic defects to explain the origin of the triangular and pentagonal cross-sections, we turned to density functional theory (DFT) and look closer at the initial epitaxial growth for further insight. The DFT surface energy calculations were performed for the surfaces illustrated in Figure 3a (exact slabs used are included in SI Figure 5) and summarised in Table 1. We observe that the {102} facets, the most stable of the observed facets, is observed in both morphologies. The {132} faces, observed in the triangular wires, and {100} facets, observed in the pseudo-pentagonal wires, both have similar surface energies, while the {112} facet has slightly higher surface energy than the other observed facets. With the surface energies we could construct a Wulff plot for the crystal shape as shown in Figure 3b. We observe that the theoretical cross section closely aligns to that observed for pseudo-pentagonal nanowires grown at 356 °C, as shown in Figure 3c. For wires grown at lower temperatures with radial overgrowth we instead saw an elongation of the {100} facets. This can be explained by the higher surface energy of the {112} facets, resulting in a comparatively higher nucleation and growth rate and subsequent elongation of the slow growing facets.[50]

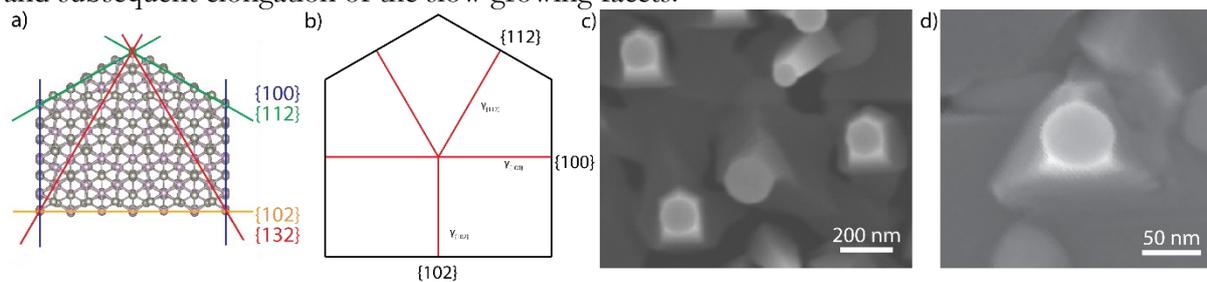

**Figure 3.** (a) Schematic cross section of the $Zn_3P_2$ crystal structure as viewed along [201]. Coloured lines indicate the facets considered in the DFT calculations. (b) Wulff plot for the pseudo-pentagonal cross section based on the DFT calculated surface energies. (c) Top view SEM image of pseudo-pentagonal nanowires grown at 356 °C with minimal radial overgrowth showing cross sections in close agreement to the calculated Wulff shape. (d) Tilted view SEM image of a $Zn_3P_2$ nanowire starting to grow from a triangular base after 5 minutes of growth.

**Table 1.** Surface energies for different $Zn_3P_2$ surfaces observed in the nanowires.

| Surface | Energy (J m$^{-2}$) | Error (J m$^{-2}$) |
| --- | --- | --- |
| {100} | 0.68 | ± 0.01 |
| {102} | 0.60 | ± 0.05 |
| {112} | 0.74 | ± 0.01 |
| {132} | 0.68 | ± 0.02 |

The Wulff plot can be used to evaluate the relative energy contributions from the different surfaces, which can then be utilised to estimate the Gibbs free energy of nucleation for both morphologies with respect to a certain base length $d$. The overall Gibbs free energy of nucleation for the triangular ($f = 3$) and pseudo-pentagonal ($f = 5$) cases can be described by the following expression:

$$\Delta G_f = (-\beta_f \Delta\mu d^2 + \alpha_f d)h \quad (1)$$

Where $\alpha_f$ is a geometry dependent term that takes into account the total surface energy per unit length of the {102} facet (*d*), $\beta_f$ is a geometry dependent term that correlates the area to the length of the {102} facet (*d*), $\Delta\mu$ is the change in chemical potential on solidification ($\Delta\mu = \Delta\mu_{vapour} - \Delta\mu_{solid}$) and *h* is the height of the nanowire. By taking the derivative and calculating the constants for each geometry (full derivation in SI), we get the following expressions to describe the nucleation barriers ($\Delta G^*$) for the different morphologies:

$$\frac{\Delta G_3^*}{h} = \frac{2.21}{\Delta\mu} \quad (2)$$

$$\frac{\Delta G_5^*}{h} = \frac{1.74}{\Delta\mu} \quad (3)$$

$\Delta\mu$ is the same for both cases, as we are investigating the same material under the same growth conditions. For a given supersaturation, the pseudo-pentagonal morphology will therefore have a lower Gibbs free energy for nucleation than the triangular case, indicating that, thermodynamically, it is the more stable phase. This is connected to the significant decrease in surface-to-volume ratio for the pseudo-pentagonal case. Thus, the triangular cross section is metastable and only achievable through some form of kinetic effect.

To explain the origin of the metastable facets we look closer at SEM images from the initial stages of growth, shown in Figure 3d. When the nanowire first starts growing it forms a triangular base related to the trifold symmetry of the Si (111) facet, which has a triangular top facet. At lower temperatures, the subsequent nanowire growth is constrained by this initial pyramid to form the metastable triangular cross section as there is not enough thermal energy to achieve the thermodynamically stable morphology. However, with increased temperature it is possible to overcome the kinetic constraints of the base and grow the more thermodynamically stable pseudo-pentagonal cross section.

Finally, the hexagonal cross section nanowires form a twin plane superlattice (TPS) morphology similar to what has been observed previously for Sn-catalysed and In-catalysed $Zn_3P_2$ nanowires.[11,41,44] The TPS nanowires are enclosed by {101} facets, which are the lowest energy facets for $Zn_3P_2$.[23] For In-catalysed VLS growth of $Zn_3P_2$ on InP (111)B wafers, the TPS morphology was observed across the whole growth parameter space when using MOCVD.[11] However, to facilitate the rotation due to the twinning in $Zn_3P_2$'s tetragonal crystal structure, there is an insertion of a monolayer of InP at the mirror plane of the twin. For Sn-catalysed $Zn_3P_2$ TPS nanowires this has to instead rely on the inclusion of Sn, which prefers to stay separate from $Zn_3P_2$ at lower temperatures.[51] Therefore, the TPS morphology is only observed at sufficiently high temperatures to allow for Sn to be incorporated into the heterotwin. However, initial measurements indicate the possibility of In contamination during these growth conditions, and further experiments to exactly determine the process will be performed.

To summarise, we report on the use of exclusively earth-abundant elements in all components for epitaxial growth of $Zn_3P_2$ nanowires using Sn as a catalyst on Si (111) substrates. Furthermore, by using Sn as a catalyst we observed various nanowire morphologies, characterised by their triangular, pseudo-pentagonal or hexagonal cross sections. The key variable controlling the morphology is temperature. Low temperatures result in the metastable triangular cross section, determined by the faceting of the initial pyramid formed during growth. By increasing the temperature, we then reach a region where the facets are no longer constrained by the base's facets, yielding the pseudo-pentagonal cross sections. Finally, at sufficiently high temperatures we observe the incorporation of Sn, allowing the formation of heterotwins and yielding the TPS morphology. These results show the wide tunability of Sn-catalysed $Zn_3P_2$ nanowire growth and demonstrate their growth on Si substrates, which is a significant step towards their potential application for sustainable solar energy harvesting. Moreover, the Sn-based heterotwins open up an alternative approach to fabricating quantum wells inside of nanowires.

# Methods

The Si (111) substrates (Siegert Wafer GmbH) were cleaned of their native oxide by dipping them in buffered oxide etchant 10:1 (Sigma-Aldrich) for 2 minutes. The samples were then rinsed using deionized water, dried using a nitrogen gun before being swiftly loaded into the Aixtron 3x2" close-coupled showerhead (CCS) MOCVD system for growth operating at a flow rate of 8 sL/min and pressure of 100 mbar. Once introduced to the MOCVD, the samples were first annealed under a phosphine ($PH_3$) atmosphere at 570 °C for 15 minutes before further heating to the particle deposition temperature of 630 °C. N.B. The temperature refers to the nominally calibrated surface temperature of the susceptor. The Sn nanoparticles were then deposited using TDMASn (Dockweiler Chemicals GmbH) as the precursor, supplied at a partial pressure of $4.97 \times 10^{-2}$ Pa for 20 minutes. The samples were then cooled to the growth temperature, which was varied from 276 °C to 396 °C. After reaching the growth temperature, a Zn pre-deposition step was carried out using DEZn as a precursor at a partial pressure of $9.72 \times 10^{-1}$ Pa for 5 minutes. To commence the growth of $Zn_3P_2$, $PH_3$ was added in addition to the DEZn at partial pressures ranging from $1 - 4.75 \times 10^1$ Pa. The growth time was 45 minutes unless otherwise specified.

Scanning electron microscopy (SEM) imaging was performed in a Zeiss Gemini operating at an acceleration voltage of 3 kV with an in-lens detector. Image analysis and particle size determination were done using the ImageJ software using the Analyze Particles function.

The focused ion beam (FIB) processed lamellae were fabricated by first transferring triangular or pseudo-pentagonal $Zn_3P_2$ nanowires to a Si dummy wafer using a micromanipulator, which in turn were inserted in the Thermo Fisher FIB Helios 5 UX system. Once in the FIB system, 0.8 μm of carbon was first deposited using the systems electron gun before subsequently depositing 2 μm of tungsten with the help of the ion gun. The tungsten helped protect the cross-sectional area in subsequent steps and improved the contrast of the images.

High-Angle Annular Dark Field (HAADF) STEM images of the cross-section of the wires were acquired in a double corrected Thermo Fisher Spectra 300 microscope operated at 300 kV. The imaging was performed with a convergence angle of 19.5 mrad and a collection angle range of 63-200 mrad. The screen current was approximately 100 pA and imaging was performed with a 1 μs dwell time with an image resolution of 4096 x 4096 pixels.

Density functional theory (DFT) calculations were performed using the Vienna *Ab initio* Simulation Package (VASP).[52,53] The exchange–correlation energy was described within the PBEsol functional[54], with a plane-wave cutoff energy of 350 eV. Brillouin-zone sampling was carried out using Γ- centred grids converged to within 0.01 eV per atom, except for geometry optimizations, which were performed with a single k-point at Γ to reduce computational cost. We verified that using Γ-only relaxations does not significantly affect the final surface energies. A vacuum spacing of 20 Å was applied along the non-periodic direction between slabs to avoid interactions between periodic images. All structures were relaxed until the residual atomic forces were smaller than 0.01 eV Å$^{-1}$. Only the ionic positions were relaxed, while the in-plane lattice constants were fixed to the bulk-optimized values.

Surface energies were evaluated for four crystallographic orientations, as stated in the main text. For each orientation, a large set of on-stoichiometric candidate slab terminations was generated to account for the multiplicity of possible atomic arrangements at the surface. All candidate slabs were first pre-relaxed using the machine-learning interatomic potential (MLIP) eSEN-30M-OAM[55], which has been shown to accurately describe systems of reduced dimensionality.[56] Because the top and bottom surfaces of a slab can undergo different reconstructions during relaxation, crystal symmetry operations (mirror or inversion, depending on orientation) were applied to enforce equivalence between the two surfaces.

From each symmetrized and relaxed configuration, a sequence of four slabs with increasing thickness (≥11 Å) was constructed by inserting additional bulk unit cells in the slab center while preserving the reconstructed surfaces. These slabs were again relaxed with eSEN-30M-OAM. A first, preliminary surface energy for each orientation and slab termination was then obtained using a slab extrapolation scheme[57,58]:

$$E_{slab}^N = 2A\gamma + Ne_{bulk} \quad (3)$$

Where $E_{slab}^N$ is the total energy of the slab, $N$ is the number of ions in the slab, $e_{bulk}$ is the bulk energy per atom, and $A$ the surface area. By applying this procedure to slabs of different thicknesses, the surface energy $\gamma$ was obtained from a linear extrapolation. For each orientation, the termination yielding the lowest surface energy was identified at the level of MLIP relaxations and subsequently re-optimized using full DFT, with final surface energies extracted through the same extrapolation procedure. The uncertainties reported in Table 1 correspond to the standard error of the linear extrapolation, obtained from regression analysis of slab energies as a function of slab thickness. Additional systematic errors associated with the choice of exchange–correlation functional, convergence parameters, and surface termination are expected to be larger, but are not included in these values.

## Acknowledgements

The authors thank Dr Mikelis Marnauza for support with preliminary TEM analysis. S. E. S. acknowledges support from the Wenner-Gren Foundation, the Åforsk Foundation and the Crafoord Foundation. The authors also acknowledge support from NanoLund, the Lund Nano Lab (myfab Lund), and Horizon Europe through the Pathfinder project SOLARUP (project number: 101046297). ICN2 acknowledges funding from Generalitat de Catalunya 2021SGR00457. ICN2 is supported by the Severo Ochoa program from Spanish MCIN / AEI (Grant No.: CEX2021-001214-S) and is funded by the CERCA Programme / Generalitat de Catalunya. Part of the present work has been performed in the framework of Universitat Autònoma de Barcelona Materials Science PhD program. Funded by the European Union as part of the Horizon Europe call HORIZON-INFRA-2021-SERV-01 under grant agreement number 101058414 and co-funded by UK Research and Innovation (UKRI) under the UK government's Horizon Europe funding guarantee (grant number 10039728) and by the Swiss State Secretariat for Education, Research and Innovation (SERI) under contract number 22.00187. Views and opinions expressed are however those of the author(s) only and do not necessarily reflect those of the European Union or the UK Science and Technology Facilities Council or the Swiss State Secretariat for Education, Research and Innovation (SERI). Neither the European Union nor the granting authorities can be held responsible for them. Authors acknowledge the use of instrumentation as well as the technical advice provided by the Joint Electron Microscopy Center at ALBA (JEMCA), in part through project 20240310035. ICN2 acknowledges funding from Grant IU16-014206 (METCAM-FIB) funded by the European Union through the European Regional Development Fund (ERDF), with the support of the Ministry of Research and Universities, Generalitat de Catalunya. ICN2 is founding member of e-DREAM.[59] The authors gratefully acknowledge the computing time provided to them on the high-performance computers Noctua 1 at the NHR Center PC2, funded by the Federal Ministry of Education and Research and the state governments participating on the basis of the resolutions of the GWK for the national high-performance computing at universities (www.nhr-verein.de/unsere-partner).

## Author Contributions

S.E.S conceived and coordinated the project. S.E.S, H.T, A.U, S.L and K.A.D performed or provided direct input to the nanowire growth experiments. N.K and S.B performed DFT calculations. F.S, M.C.S and J.A performed FIB and STEM analysis. S.E.S and J.J derived the thermodynamic model. S.E.S wrote the manuscript with input from all co-authors.

## Competing Interests

The authors declare that they have no competing interests.

# Supplementary information to "Unconventional cross sections in zinc phosphide nanowires grown using exclusively earth-abundant components"

Simon Escobar Steinvall[1*], Hampus Thulin[1], Nico Kawashima[2,3], Francesco Salutari[4], Jonas Johansson[5], Aidas Urbonavicius[1], Sebastian Lehmann[1], Maria Chiara Spadaro[4,6], Jordi Arbiol[4,7], Silvana Botti[2], Kimberly A. Dick[1]

1. Center for Analysis and Synthesis and NanoLund, Lund University, Box 124, 221 00 Lund, Sweden
2. Research Center Future Energy Materials and Systems and Interdisciplinary Centre for Advanced Materials Simulation, Faculty of Physics and Astronomy, Ruhr University Bochum, Universitätsstraße 150, 44801 Bochum, Germany
3. Institute of Condensed Matter Theory and Optics, Friedrich-Schiller-Universität Jena, Max-Wien-Platz 1, 07743 Jena, Germany
4. Catalan Institute of Nanoscience and Nanotechnology (ICN2), CSIC and BIST, Campus UAB, Bellaterra, 08193 Barcelona, Catalonia, Spain
5. Division of Solid State Physics and NanoLund, Lund University, Box 124, 221 00 Lund, Sweden
6. Department of Physics and Astronomy "Ettore Majorana", University of Catania and CNR-IMM Via S. Sofia 64, 95123 Catania, Italy
7. ICREA, Pg. Lluís Companys 23, 08010 Barcelona, Catalonia, Spain
*Corresponding Author: simon.escobar_steinvall@chem.lu.se

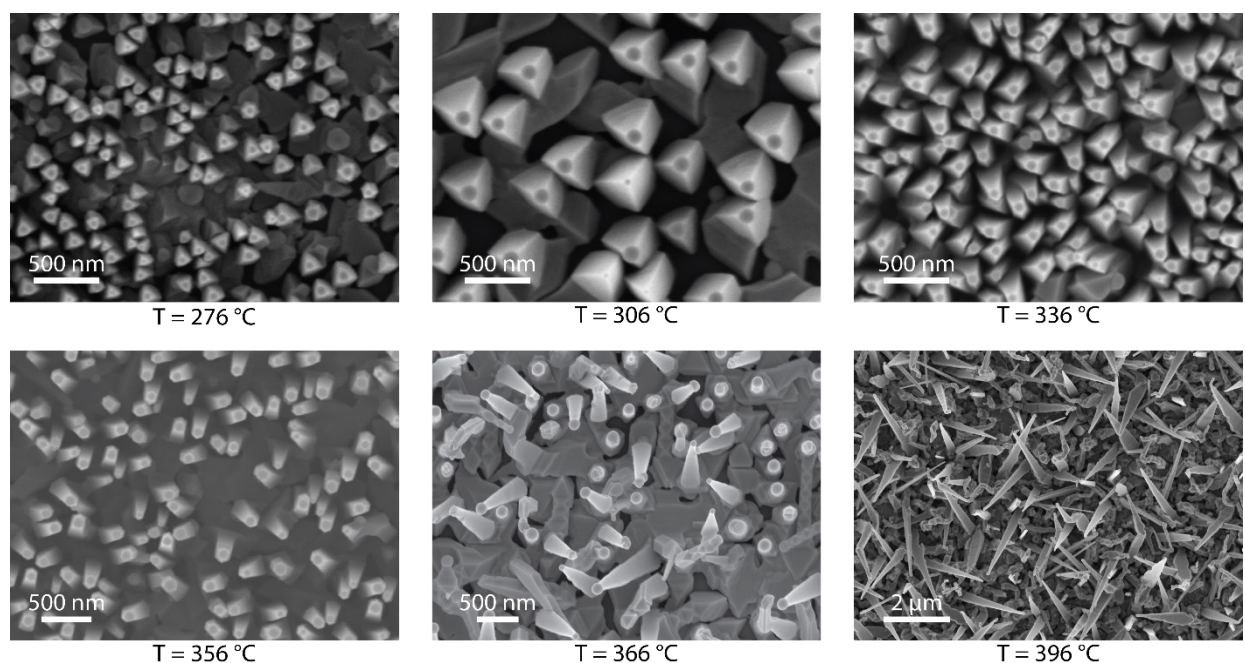

**SI Figure 1.** Top view SEM images of Sn-catalysed $Zn_3P_2$ nanowires grown on Si (111) with a V/II ratio of 36.0 for 45 minutes at varying temperatures.

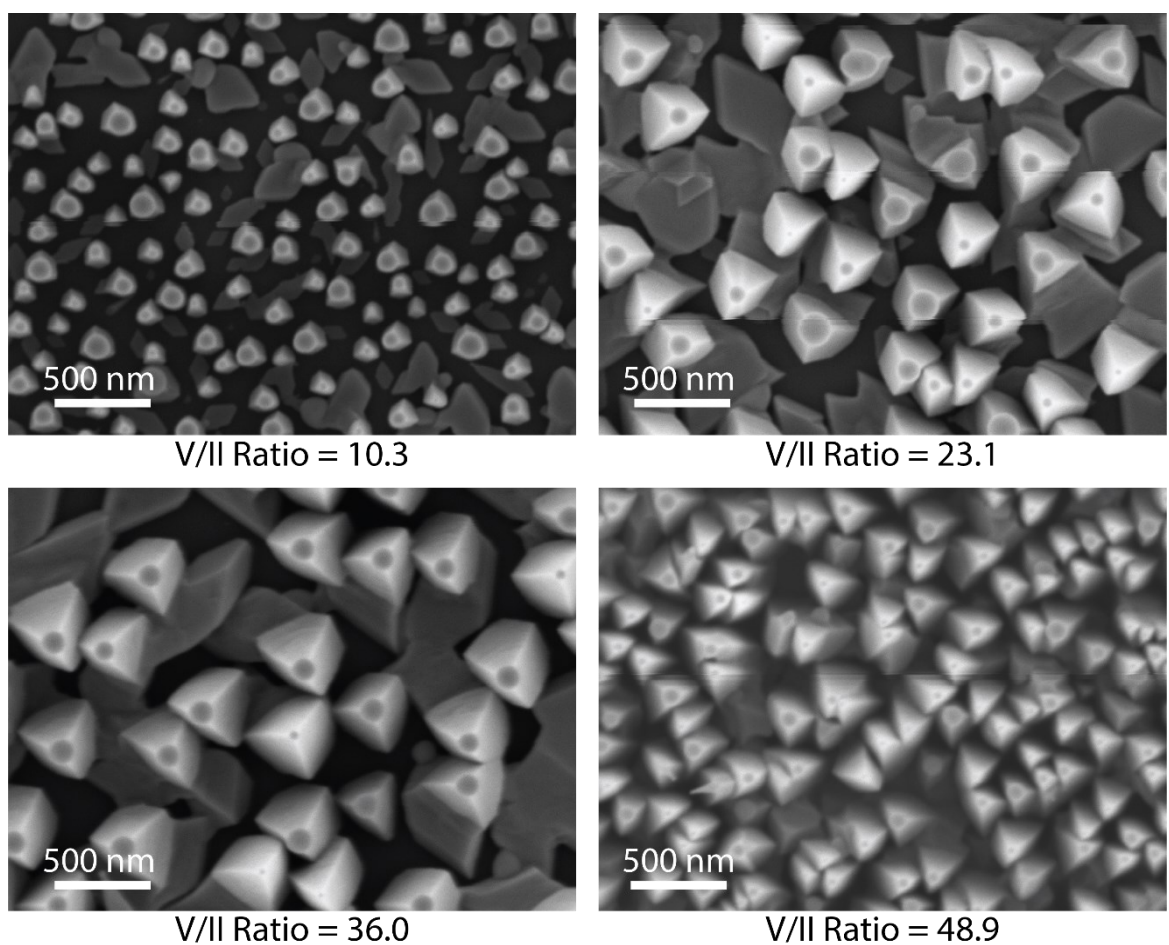

**SI Figure 2.** Top view SEM images of Sn-catalysed $Zn_3P_2$ nanowires grown on Si (111) at 306 °C for 45 minutes under varying $PH_3$ partial pressures.

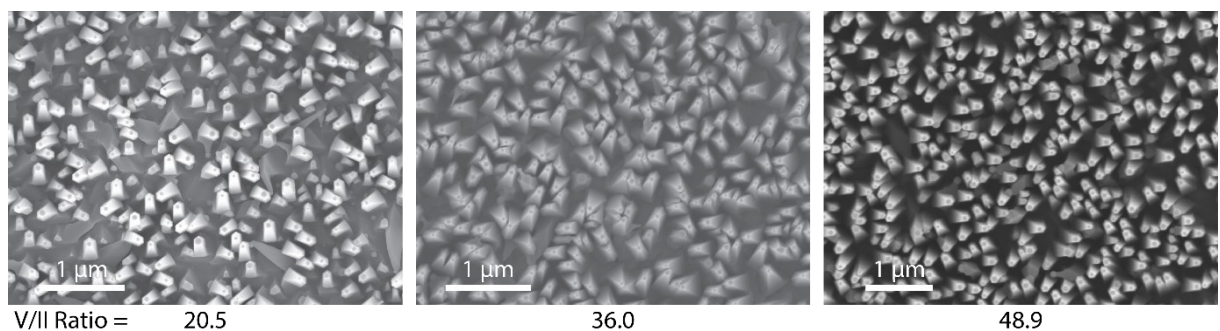

**SI Figure 3.** Top view SEM images of Sn-catalysed $Zn_3P_2$ nanowires grown on Si (111) at 336 °C for 45 minutes under varying $PH_3$ partial pressures.

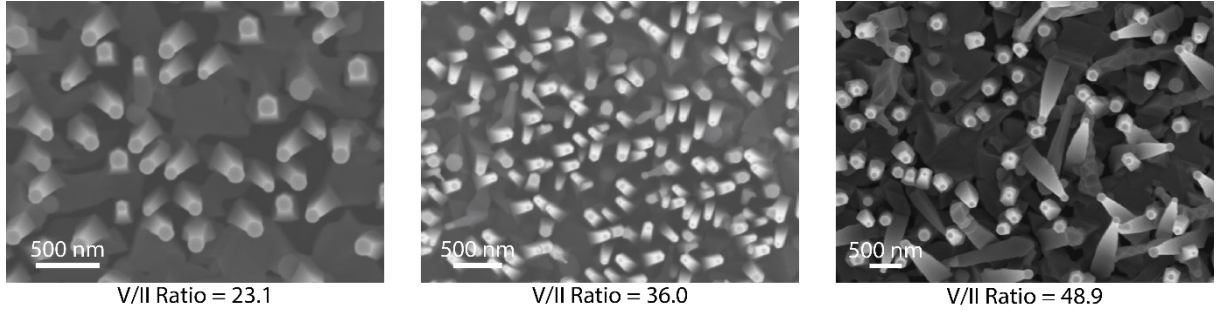

| V/II Ratio = 23.1 | V/II Ratio = 36.0 | V/II Ratio = 48.9 |

**SI Figure 4.** Top view SEM images of Sn-catalysed $Zn_3P_2$ nanowires grown on Si (111) at 356 °C for 45 minutes under varying $PH_3$ partial pressures.

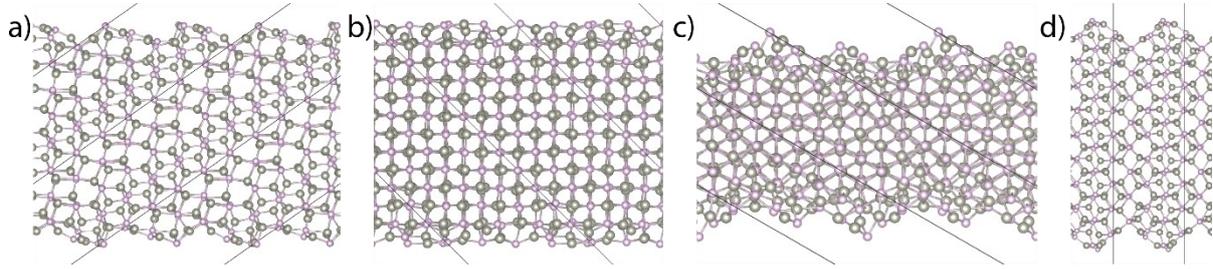

**SI Figure 5.** Slabs used for the DFT calculations, shown after full geometry optimisation, indicating the cut location and atomic structure of (a) the {102} as viewed along the [010] direction, (b) {112} as viewed along the [110] direction, (c) {132} as viewed along the [-60-3] direction, and (d) {100} as viewed along the [010] direction.

**Gibb's free energy of nucleation derivation**

The Gibb's free energy of the system will be described by the sum of the surface and the bulk contributions, simplified to a two-dimensional cross section. For a base length $d$ (length of {102} in both cases), the energy per unit of cross section with $f$ number of facets can be described by the following expressions:

$$\Delta G_f = (-\beta_f \Delta \mu d^2 + \alpha_f d)h \quad (S1)$$

Where $\alpha_f$ is a term that takes into account the total surface energy per unit length of the {102} facet (*d*), $\beta_f$ is a geometry dependent term that correlates the area to the length of the {102} facet (*d*), $\Delta \mu$ is the change in chemical potential on solidification ($\Delta \mu = \Delta \mu_{vapour} - \Delta \mu_{solid}$) and *h* is the height of the nanowire.

For $f = 3$, the triangular cross section nanowires, we have an equilateral triangle. For this case α becomes:

$$\alpha_3 = \gamma_{\{102\}} + 2\gamma_{\{132\}} = 1.952\, J\, m^{-2} \quad (S2)$$

Moreover, for an equilateral triangle $\beta_3 = \sqrt{3}/4$.

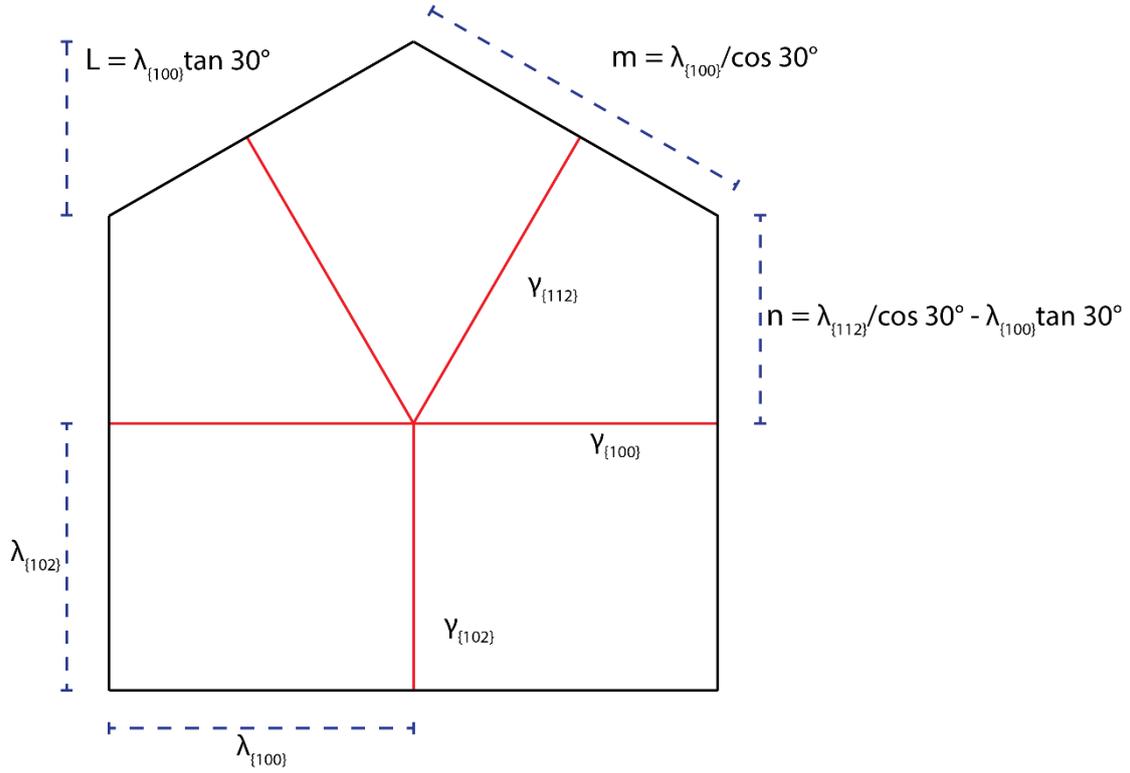

**SI Figure 6.** Wulff construct with additional information on distances used in the calculations.

For $f = 5$, the pseudo-pentagonal case it is less straight forward. Based on the Wulff construct in SI Figure 6 we can get the following relationship between the different sides as a function of $d$:

$$2\lambda_{\{100\}} = d \quad (S3)$$

$$\lambda_{\{102\}} = \frac{\gamma_{\{102\}}}{\gamma_{\{100\}}} \lambda_{\{100\}} = \frac{\gamma_{\{102\}}}{2\gamma_{\{100\}}} d \quad (S4)$$

$$\lambda_{\{112\}} = \frac{\gamma_{\{112\}}}{\gamma_{\{100\}}} \lambda_{\{100\}} = \frac{\gamma_{\{112\}}}{2\gamma_{\{100\}}} d \quad (S5)$$

Using this we first derive $\beta_5$ through the following expression:

$$\beta_5 d^2 = (\lambda_{\{102\}} + n) 2\lambda_{\{100\}} + 2\lambda_{\{100\}} \frac{\lambda_{\{100\}} \tan 30°}{2} =$$

$$= \left(\lambda_{\{102\}} + \frac{\lambda_{\{112\}}}{\cos 30°} - \lambda_{\{100\}} \tan 30°\right) 2\lambda_{\{100\}} + 2\lambda_{\{100\}} \frac{\lambda_{\{100\}} \tan 30°}{2} =$$

$$= 2\lambda_{\{100\}} \left(\lambda_{\{102\}} + \frac{\lambda_{\{112\}}}{\cos 30°} - \lambda_{\{100\}} \tan 30°\right) =$$

$$= d \left(\frac{\gamma_{\{102\}}}{2\gamma_{\{100\}}} d + \frac{\gamma_{\{112\}}}{2\gamma_{\{100\}}} \frac{1}{\cos 30°} d - \frac{d}{2} \frac{\tan 30°}{2}\right) =$$

$$= \frac{1}{2} \left(\frac{\gamma_{\{102\}}}{\gamma_{\{100\}}} + \frac{\gamma_{\{112\}}}{\gamma_{\{100\}}} \frac{1}{\cos 30°} - \frac{\tan 30°}{2}\right) d^2$$

Which gives:
$$\beta_5 = 0.93$$

We can then derive $\alpha_5$ as follows:

$$\alpha_5 d = 2\lambda_{\{100\}}\gamma_{\{102\}} + 2(\lambda_{\{102\}} + n)\gamma_{\{100\}} + 2m\gamma_{\{112\}} =$$

$$= \gamma_{\{102\}}d + 2\left(\frac{\gamma_{\{102\}}}{2\gamma_{\{100\}}}d + \frac{\gamma_{\{112\}}}{2\gamma_{\{100\}}}\frac{d}{\cos 30°} - \frac{d}{2}\tan 30°\right)\gamma_{\{100\}} + 2\frac{d}{2}\frac{1}{\cos 30°}\gamma_{\{112\}} =$$

$$= d\left(\gamma_{\{102\}} + \gamma_{\{102\}} + \frac{\gamma_{\{112\}}}{\cos 30°} - \gamma_{\{100\}}\tan 30° + \frac{\gamma_{\{112\}}}{\cos 30°}\right) =$$

$$= d(2\gamma_{\{102\}} + \frac{2}{\cos 30°}\gamma_{\{112\}} - \gamma_{\{100\}}\tan 30°)$$

Which gives:

$$\alpha_5 = 2.52\ J\ m^{-2}$$

The Gibb's free energy of nucleation based on equation S1 for the triangular and pseudo-pentagonal can thus be described by the following equations:

$$\Delta G_3 = 1.95hd - 0.43\Delta\mu hd^2 \quad (S6)$$

$$\Delta G_5 = 2.52hd - 1.14\Delta\mu hd^2 \quad (S7)$$

To estimate the critical radius for nucleation we first take the derivative of S1 with respect to $d$:

$$\Delta G'_f = \alpha_f h - 2\beta_f \Delta\mu hd \quad (S8)$$

And to get the critical base length for nucleation, $d^*$ we look at the maximum, which is when $\Delta G'_3 = 0$, giving

$$d^* = \frac{\alpha_f}{2\beta_f \Delta\mu} \quad (S9)$$

If we plug this back into (S1) and make it per unit height we get:

$$\frac{\Delta G^*_f}{h} = \frac{\alpha_f^2}{2\beta_f \Delta\mu} - \frac{\alpha_f^2 \beta_f \Delta\mu}{(2\beta_f \Delta\mu)^2} = \frac{\alpha_f^2}{4\beta\Delta\mu} \quad (S10)$$

By plugging in the values for $f=3$ and $f=5$ we get the following equations:

$$\frac{\Delta G^*_3}{h} = \frac{2.21}{\Delta\mu} \quad (S11)$$

$$\frac{\Delta G^*_5}{h} = \frac{1.74}{\Delta\mu} \quad (S12)$$

As 2.21>1.74 the pseudo-pentagonal case will have a lower Gibb's free energy of nucleation for all growth conditions, as the chemical potential will be the same as it is the same material and growth conditions analysed in both cases.